\documentclass[preprint]{aastex62}

\submitjournal{AAS Journals}

\shorttitle{Extinction in Dippers}
\shortauthors{Sitko et al.}
\begin{document}

\title{Wavelength-Dependent Extinction and Grain Sizes in ``Dippers''}

\correspondingauthor{Michael L Sitko}
\email{michael.sitko@uc.edu}

\author[0000-0003-1799-1755]{Michael L. Sitko}
\affil{Center for Exoplanetary Systems, Space Science Institute, 4750 Walnut Street, Suite 205, Boulder, CO 80301, USA}

\author[/0000-0002-7818-2305]{Ray W. Russell}
\affil{The Aerospace Corporation, Los Angeles, CA, USA}

\author{Zachary C. Long, Korash Assani, Monika Pikhartova}
\affiliation{Department of Physics, University of Cincinnati, Cincinnati, OH 45221-0011, USA}

\author{Ammar Bayyari}
\affiliation{Department of Physics, University of Cincinnati, Cincinnati, OH 45221-0011, USA}
\affiliation{Department of Physics and Astronomy, University of Hawaii, 2505 Correa Road Watanabe 416. Honolulu, Hawaii 96822 USA}

\author{Carol A. Grady}
\affiliation{Eureka Scientific, Inc., 2452 Delmer Street, Suite 100, Oakland, CA 496002, USA}

\author[0000-0002-9548-1526]{Carey M. Lisse}
\affiliation{JHU-APL, 11100 Johns Hopkins Road, Laurel, MD 20723, USA}

\author[0000-0001-9910-9230]{Massimo Marengo}
\affiliation{Department of Physics and Astronomy, Iowa State University, Ames, IA 50010, USA}

\author[0000-0001-9209-1808]{John P. Wisniewski}
\affiliation{Department of Physics and Astronomy, George Mason University, 4400 University Drive, Fairfax, VA 22030-4444, USA}

\author[0000-0002-9209-5830]{William Danchi}
\affiliation{NASA Goddard Space Flight Center, 8800 Greenbelt Road, Greenbelt, MD 20771-2400, USA}

%% Note that the \and command from previous versions of AASTeX is now
%% depreciated in this version as it is no longer necessary. AASTeX 
%% automatically takes care of all commas and "and"s between authors names.

%% AASTeX 6.2 has the new \collaboration and \nocollaboration commands to
%% provide the collaboration status of a group of authors. These commands 
%% can be used either before or after the list of corresponding authors. The
%% argument for \collaboration is the collaboration identifier. Authors are
%% encouraged to surround collaboration identifiers with ()s. The 
%% \nocollaboration command takes no argument and exists to indicate that
%% the nearby authors are not part of surrounding collaborations.

%% Mark off the abstract in the ``abstract'' environment. 
\begin{abstract}
We have examined inter-night variability of K2-discovered ``Dippers" that are not close to being viewed edge-on, as determined from previously-reported ALMA images, using the SpeX spectrograph on NASA's Infrared Telescope facility (IRTF). The three objects observed were EPIC 203850058, EPIC 205151387, and EPIC 204638512 ( = 2MASS J16042165-2130284). Using the ratio of the fluxes between two successive nights, we find that for EPIC 204638512 and EPIC 205151387, we find that the properties of the dust differ from that seen in the diffuse interstellar medium and denser molecular clouds. However, the grain properties needed to explain the extinction does resemble those used to model the disks of many young stellar objects. The wavelength-dependent extinction models of both  EPIC 204638512 and EPIC 205151387 includes grains at least 500 microns in size, but lacks grains smaller than 0.25 microns. The change in extinction during the dips, and the timescale for these variations to occur, imply obscuration by the surface layers of the inner disks. The recent discovery of a highly mis-inclined inner disk in EPIC 204638512 is suggests that the variations in this disk system may point to due to rapid changes in obscuration by the surface layers of its inner disk, and that other ``face-on'' Dippers might have similar geometries. The He I line at 1.083 microns in EPIC 205151387 and EPIC 20463851 were seen to change from night to night, suggesting that we are seeing He I gas mixed in with the surface dust. 
\end{abstract}


\section{Introduction} \label{sec:intro}

The formation of planets in young protoplanetary disk systems by means of core accretion requires that the dust within these systems undergo growth and likely settling of the largest grains toward the disk midplane. The degree of such grain growth, however, is often difficult to assess observationally. At visible- to mid-infrared wavelengths, the disks may be optically thick, and observations only give clues about the surface layers. Both the scattered light and thermal emission are subject to the effects of radiative transfer, and their interpretation will be model-dependent. At millimeter/submillimeter wavelengths, the spectral slope provide clues as to grain sizes, but optical thickness and grain size distributions will not be constant throughout the planet-forming regions of the disk. In particular,  gas/dust mass ratios derived from such grains seem  to unusually low (by factors of up to 30 times compared to expected gas/dust ratios) \citep{bernstiel10,williams14,kama16,long17}.  Recent models of the depletion of CO in young disk systems suggests that gaseous CO will condense into ices, and coat more refractory grains, and the extent of the depletion depends on radial distance and age of the system \citep{mcclure16,powell22}. Mid-IR interferometry near 10 $\mu$m indicates that grains are often larger near the star than further out \citep{vB04,lopez22}. \\

Occultations of the stellar light by dust along the line of sight, however, can provide a more direct measure of the wavelength-dependent extinction of the dust in these systems at visible and near-IR wavelengths. Many young stars undergo variations in brightness that are most easily explained by dust extinction in the upper layers of disks that are viewed at high inclinations. Such ``UX Orionis'' stars, or ``UXors'' have been known for some time \citep{grinin91,grinin96,eaton_herbst95}. In the case of disks with puffed-up inner rims \citep{natta01,dd04} the occulting dust would likely be located at the inner disk rim of highly self-shadowed disks \citep{dullemond03}. The color changes in these systems have been studied at visible wavelengths, and as expected for submicron-sized grains, the extinction is accompanied by reddening, until the dust becomes optically thick, after which "bluing" may actually occur due to the remaining stellar light reaching the observer relatively unaffected. This occurs when the disk is very optically thick, but scattering at the poles reaches the observer \citep{grinin91}.  \\

The \textit{Spitzer} Infrared Array Camera (IRAC) and the Convection, Rotation, and Planetary Transit (COROT) satellite provided long time baseline photometric observations of large numbers of Classical T Tauri stars (CTTs) in young stellar clusters. Among these were stars exhibiting large irregular changes in brightness over time scales of days or even hours. These ``Dippers'' became important targets for the K2 mission, and  \citet{ansdell16a} reported finding many in the Upper Sco and $\rho$ Oph stellar associations.  Similar ``dipper" activity was observed in stars in the young star-formation region NGC 2264 \citep{stauffer16}. While disks with near edge-on orientations would would be most likely capable of occulting the light of the star, \citet{ansdell16b} presented ALMA continuum observations of three Dippers, EPIC 203850058, EPIC 205151387, and EPIC 204638512, whose geometries were far from edge-on. In the case of EPIC 204638512, the disk is nearly face-on ($\sim 6 \pm 1.5^\circ$.), with a large dust cavity, yet it exhibits brightness fluctuations of nearly 1 magnitude over a time span of a day. This suggests that if the brightness changes were due to dust, a UXor-type geometry was unlikely. \\

However, recent imaging data for EPIC 204638512  ALMA \citep{mayama18} and VLT/SPHERE \citep{pinilla18} provide evidence for an inner disk that is misaligned with the outer disk. With ALMA observations, a twisted  ``butterfly'' radial velocity map in CO (3-2) clearly shows a highly inclined inner disk, despite the outer disk being seen nearly face-on. Shadows of the inner disk cast on the outer disk are also visible in both dust continuum thermal emission and gas emission maps. These shadows are also visible in the scattered light images from SPHERE, where the depth and width of the shadows change with time. In the case of this Dipper, at least, the light variations are likely coming from irregularities in the orbiting dust in the inner disk. In the other two objects discussed in this paper, we need more evidence to conclude a similar geometry, because there are other ways to generate similar variability in brightness. \\

Other possible mechanisms may also produce somewhat similar changes in the observed brightness of a star. Using high spatial resolution imaging obtained with the Goddard Fabry-P\'{e}rot  imager on the Apache Point Observatory 3.5 m telescope, \citet{was06} reported a string of luminous clouds along two opposing outflows in the Herbig Ae star HD 163296 (MWC 275). \citet{ellerbroek14}, using photometry obtained by the All Sky Automated Survey \citep{pojmanski97}, reported that a Dipper-like extinction event occurred in 2001 in the star, and this extinction event was traced, using the measured proper motion of these clouds, to an ejection event at that time. it was suggested that dust entrained in a disk wind (see their Figure 9), similar to that suggested by \citet{bans12} for the source of the near-IR emission in protostellar systems, was the cause. Unfortunately, the ASAS photometric observations provided no color information. However, in 2012 multi-filter observations obtained by the American Association of Variable Star Observers (AAVSO) captured a 1-day extinction event that could be subjected to modeling using various types of grains. Grains typical of both the diffuse interstellar medium and dark molecular clouds with star-formation regions provided poor fits to the wavelength dependence of the extinction. However grains that had undergone further growth, and were similar to those included in some radiative transfer models of disks, such as those used in the HOCHUNK3D code package by \citet{whitney13}, were more successful at reproducing the extinction properties of that one-day event \citet{pikhartova21}  \\

Here we report near-infrared observations of the three Dippers in \citet{ansdell16b} obtained with the SpeX spectrograph on NASA's Infrared Telescope facility (IRTF) taken on two consecutive nights. From these we obtain the extinction curves for the two brighter ``Dippers", and subject them to the same extinction analysis that \citet{pikhartova21} employed for HD 163296. \\

\section{Observations}

We observed the three ``face-on'' Dippers on two successive nights, 10 and 11 August, 2017, at an effective spectral resolution R = 750 (0.8" slit) from 0.7 - 2.4 $\mu$m using the SXD grating of the SpeX spectrograph \citep{rayner03} on NASA's Infrared Telescope Facility (IRTF). In order to determine the absolute fluxes of these slit spectra, which are subject to time-variable throughput due to the effects of astronomical seeing and telescope guiding, we also observed them in the K-band using the SpeX guide camera using a 9-point dither pattern. Both spectra and images were flux-calibrated using the A0V star HD 145127 at nearly the same airmass. The spectral data reduction was carried out using the Spextool software, running under IDL. \citep{vacca03,cushing04} \\

\begin{figure}
\plotone{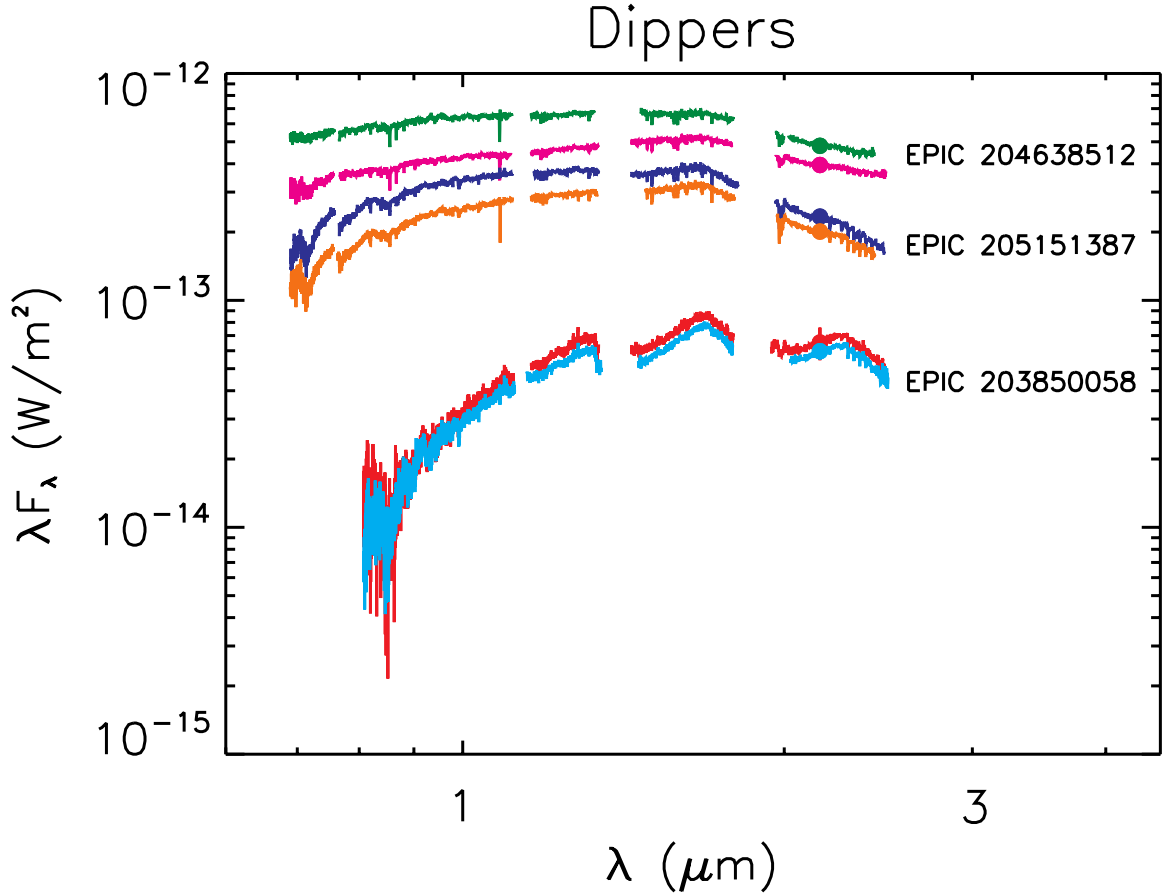}
\caption{0.7-2.4 $\mu$m spectra of three Dippers obtained with the SpeX spectrograph between 0.7-2.4 $\mu$m on 10 and 11 Aug. 2017 (UT). Due to poor signal/noise, the spectrum of EPIC 203850058 was truncated at wavelengths smaller than 0.8 $\mu$m. The spectra were calibrated in absolute flux using K-band images obtained with the guide camera (filled circles). For EPIC 204628512 and EPIC 205151387, the fainter spectrum was also slightly redder, and were used to examine the wavelength-dependence of the extinction (reddening). The difference in the flux levels for EPIC 2035850058 were so small that we did not attempt to derive a flux ratio for examining potential reddening by dust. \label{fig:f1}}
\end{figure}

In Figure 1 we show the SpeX spectra of all three Dippers, each normalized by the K-band fluxes. The two brightest Dippers,  EPIC 204638512 (K2; \citet{kohler00}) and EPIC 205151387 (M0.1V; \citet{ansdell16a}), underwent marked changes in brightness between the two nights, while the third, much fainter one, EPIC 203850058 (M5; \citet{manora15}), showed smaller changes, being less than the ``noise" at the shortest wavelengths. \\

\begin{deluxetable}{lccc}
\tablecolumns{7}
\tablewidth{0pc}
%\rotate
\tabletypesize{\scriptsize}
\tablecaption{Dipper Fundamental Parameters}
\tablehead{
\colhead{Star Name} & \colhead{EPIC 203850058}  & \colhead{EPIC 204638512} & \colhead{EPIC 205151387}  }

\startdata
\emph{Dippers} & & &  \\
2MASS & 16253849-2613540  & 016042165-2130284  & 16090075-1908526    \\
Gaia DR3 & 6049072316675781120 & 6243393817024157184 & 6245777283349430912  \\
\emph{Star Properties} & & & \\
Gaia Distance (pc) & 141.577 & 145.309 & 137.397 \\
Spectral Type & M5 & K2 & M0.5 \\
P$_{rot}$ (d) & 2.88 & 5.0 & 9.55 \\
Region & $\rho$ Oph & Upper Sco & Upper Sco \\
K Band Magnitude & 10.766 & 8.508 & 9.475 \\
\emph{Disk Properties} & & & \\
ALMA Disk Diam. (") & 0.2 x 0.3 & 1 & 0.6 x 0.4 \\
ALMA Disk Diam. (au) & 28.3 x 42.5 & 145 & 82.2 x 54.8 \\
AlLMA Disk Incl. ($^\circ$) & 73$\pm$23 & 6$\pm$1.5 & 53$\pm$2 \\
Dusk Type \tablenotemark{a} & Full & Transitional & Full \\
Age & 1 Myr & 10 Myr & 10 Myr \\
\enddata
\tablenotetext{a}{\citep{ansdell16a}}
\end{deluxetable}

\begin{deluxetable}{ccccccc}
\tablecolumns{7}
\tablewidth{0pc}
%\rotate
\tabletypesize{\scriptsize}
\tablecaption{Observed Dipper K-Band Fluxes\tablenotemark{a}}
\tablehead{
\colhead{Date (UT)} & \colhead{MJD}  & \colhead{EPIC 203850058} & \colhead{MJD}  & \colhead{EPIC 204638512} & \colhead{MJD} & \colhead{EPIC 205151387}  }

\startdata
170810 & 57975.23912 & 0.65(0.11) & 57975.25421 &  3.95(0.12) & 57975.27806 &  2.34(0.11)   \\
170811 & 57976.24319 & 0.60(0.05) & 57976.24319 & 4.80(0.04) & 57976.27624 & 2.01(0.04)  \\
\enddata
\tablenotetext{a}{ ~K-band flux is $\lambda$F$_{\lambda}$in units of 10$^{-13}$ W m$^{-2}$. Uncertainties are given in parentheses.}
\end{deluxetable}

\section{Wavelength-Dependent Extinction}

To characterize the dust we first selected the brightest of these objects, EPIC 204638512 ( = 2MASS J16042165-2130284), an object that has been the target of imaging programs using extreme adaptive optics \citep{dong17,uyama16,canovas17}.  The difference in extinction between the two successive nights (we cannot assume that the brighter flux state has no extinction) was determined by taking the ratio of the two flux-calibrated spectra, and comparing the result with a number of published dust extinction laws, as shown in Figure 2. \\

\begin{figure}
\plotone{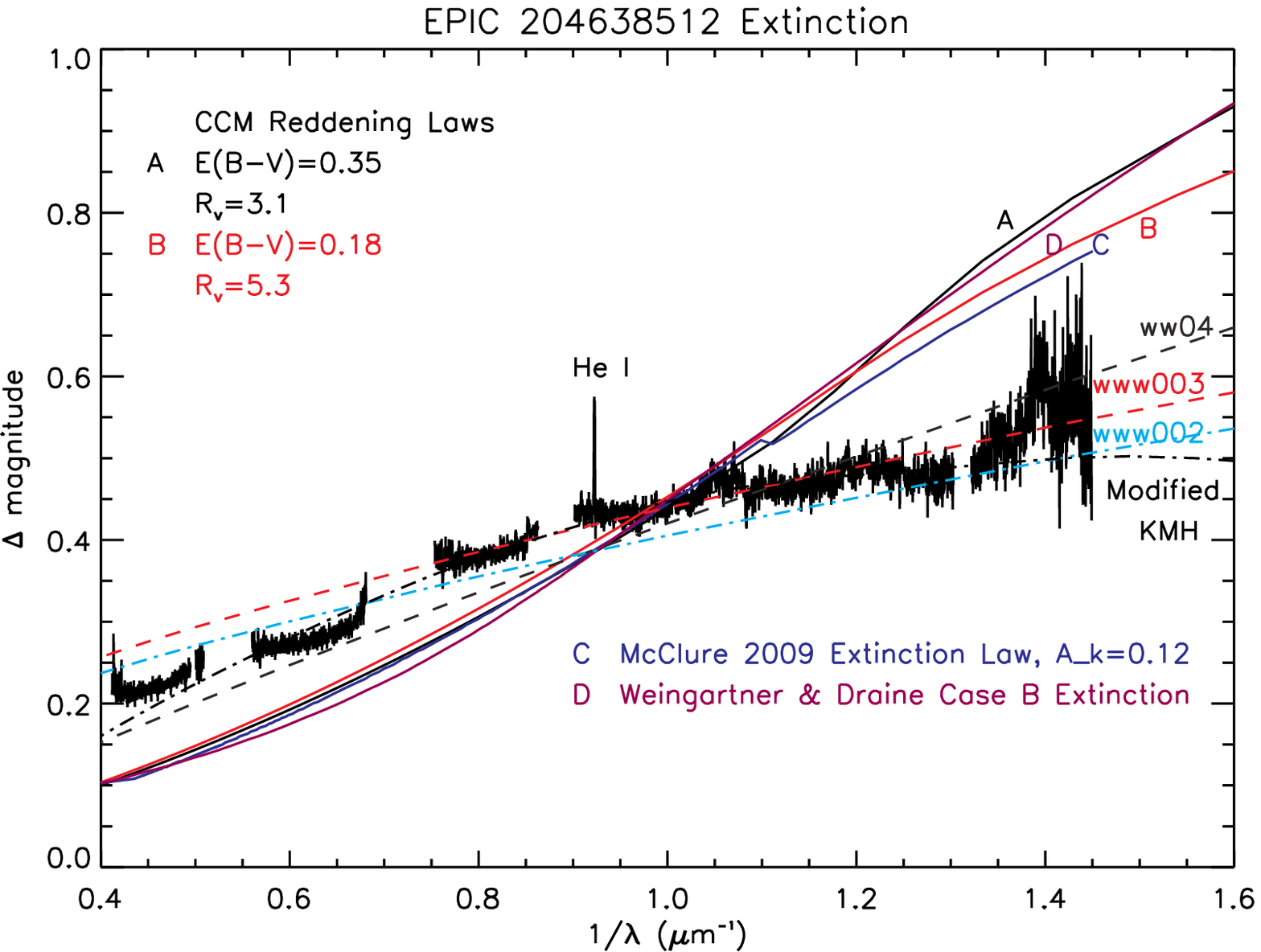}
\caption{Extinction curve of  EPIC 204638512. The model curves labeled ``CCM''. are those of \citet{ccm89}, using different values of R$_{V}$, and representative of ISM extinction. Also shown are DMC extinction curves from  \citet{mcclure09}  and \citet{wd01} The curves labeled ``ww04'', ``www003'', and ``ww002'' are grain files included with the HOCHUNK3D code package of \citet{whitney13}. For example, ``www003''' is Model 1 of \citet{wood02}, which has a power-law grain size distribution, with a maximum grain size of 1000 $\mu$m and an exponential cutoff at starting an 50 $\mu$m and going to the large grain limit. ``Modified KMH'' uses the grain size distribution of \citet{kmh94} but increasing the minimum and maximum grain sizes to 0.25 $\mu$m and 500 $\mu$m, respectively. These are described more fully in the text. \label{fig:f2}}
\end{figure}

\begin{figure}
\plotone{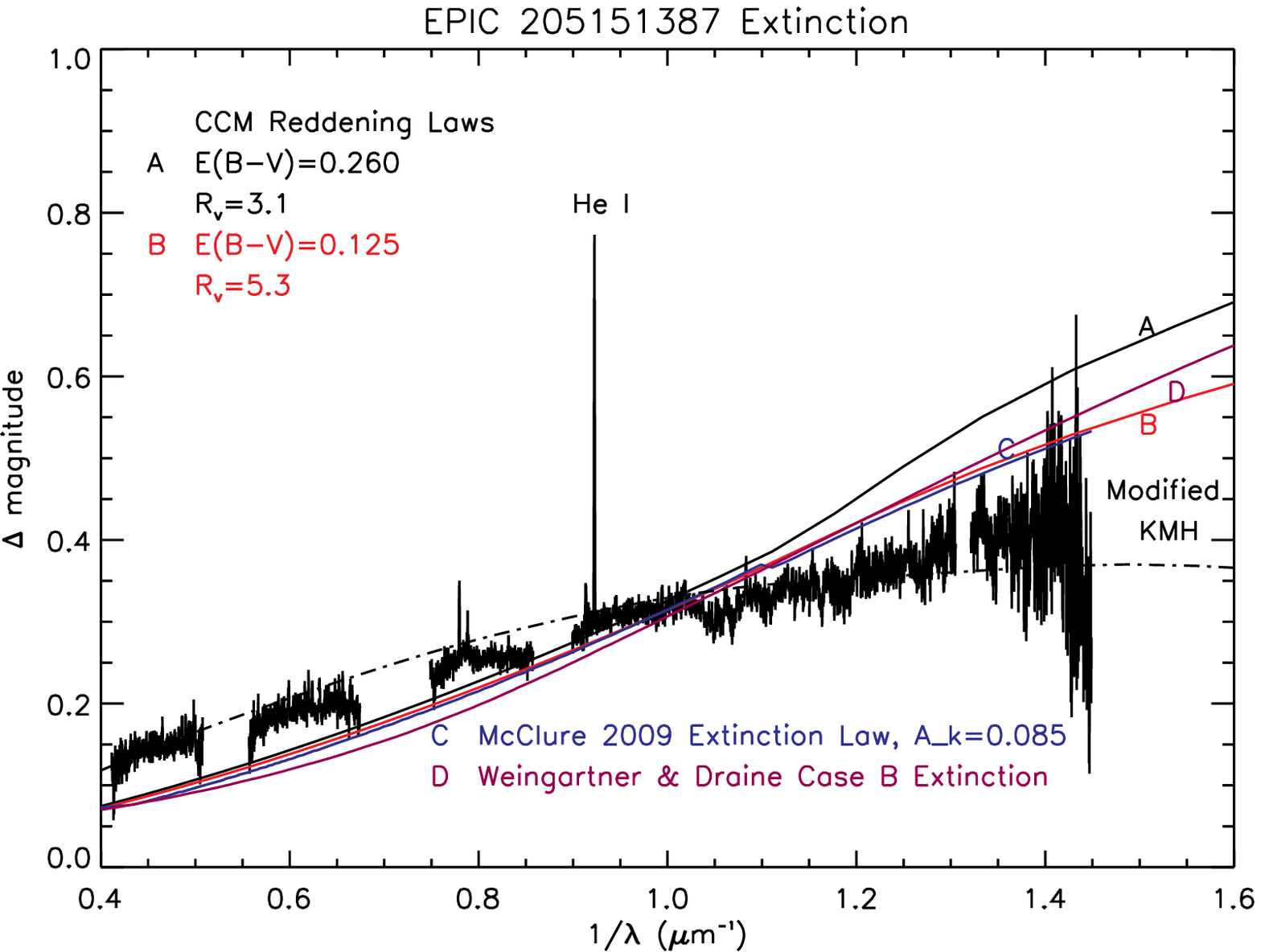}
\caption{Extinction curve of EPIC 205151387. The grain models are explained in the caption to Figure 2, except we did not include grain models used in the HOCHUNK3D package, however it is clear that the grain sizes involved are similar to those in EPIC 204638512. The change in the He I feature is more pronounced in EPIC 205151387. \label{fig:f3}}
\end{figure}

We began with the extinction laws of \citet{ccm89} often used in the literature for reproducing extinction by dust in the diffuse interstellar medium. Despite changing the ratio of total-to-selective extinction (R$_{V}$) where larger values represent larger grains), no good fit was obtained. Other recipes for denser cloud regions \citep{wd01,mcclure09}  fared little better. \\

Next we tried grain prescriptions (e.g. from \citet{cotera01} and \citet{wood02}) developed for circumstellar disks and included in the HOCHUNK3D Monte Carlo radiative transfer code of  \citet{whitney13}. These use the optical constants of silicates and carbonaceous grains from  \citet{rouleau91}  and \citet{wd01}. ``www03" refers to the grain created to the disk midplane providing the bet fit for the SED of HH30 in  \citet{wood02}. Specifically, their Model 1, which uses a mixture of amorphous carbon and astronomical silicates that have a power-law size distribution $n(a)\sim a^{-p}$ with $p= 3.0$, plus exponential cutoff starting at 50$\mu$m and going to longer wavelengths, to a grain maximum size size of 1000 $\mu$m. ``ww04" is the grain model of  \citet{cotera01}, used to fit the scattered light in HH30. It a power law index of 3.5 for amorphous carbon and 3.0 for silicates, a turnover at 0.55 $\mu$m and grain maximum of 20 $\mu$m. ``ww002" is the same grain model as ``w003" except having the exponential cutoff at 100 $\mu$m instead of 50 $\mu$m. \\

Finally we used a similar grain file that used the ISM dust extinction of \citep{kmh94}  but adjusted the minimum and maximum size present. This extinction (``Modified KMH'') has grains extending to at least 500 $\mu$m and lacks grains smaller than 0.25 $\mu$m. While larger grains were not excluded, they produced little further change in the resulting extinction at the observed wavelengths. \\

We also subjected the extinction curve of EPIC 205151387 to a similar analysis, as illustrated in Figure 3. This uses the same modified KMH model as in the case of EPIC 204638512 (with lower total extinction), grains more consistent with those used in disk modeling provided a much better agreement with the observed extinction than did either ISM-like or dark cloud-like extinction models. \\

A similar analysis was applied to a short-duration extinction event in HD 163296 by \citet{pikhartova21}. In the case of HD 163296, an extinction event lasting only about 1 day was modeled, with the www003 grain model providing the best fit, over all. The large extinction event of 2001 \citep{ellerbroek14}, likely due to the ejection of HH-like blobs in a bipolar jet (with a disk wind) was mostly monitored at a single visible wavelength, and could not be subjected to the same wavelength-dependent extinction analysis.\\

\section{The Gas}

In both derived extinction curves, there is an added feature at a wavelength consistent with the He I line at 1.083 $\mu$m often seen in other young stellar objects. In EPIC 204638512,  this line seems to exhibit exhibits a ``P Cygni" profile (blueshifted absorption component flanked by a redshifted emission component) indicative of outflowing gas (with the blue absorption wing being produced by gas between the observer and the stellar photosphere), but the evidence is only suggestive. \citet{sa20} reported  line profile changes in H$\alpha$ and H$\beta$. The former line exhibited complex behavior, with epochs of inflowing and outflowing gas, while the latter had one epoch with a very distindegreect inflowing gas component  In EPIC 205251387, the He I line is more indicative of infalling gas. In PMS stars, the He I profile  can exhibit  strong time-dependent changes in intensity and shape  \citep{sitko12,fernandes18}. A significant fraction of the time, HD 163296 seems to exhibit a combination of both outflowing and inflowing gas \citep{pikhartova21} at the same time.  The data on EPIC 203850058 was deemed to ne of insufficient quality to determine the nature of the gas flow characteristics. \\

\begin{figure}
\plotone{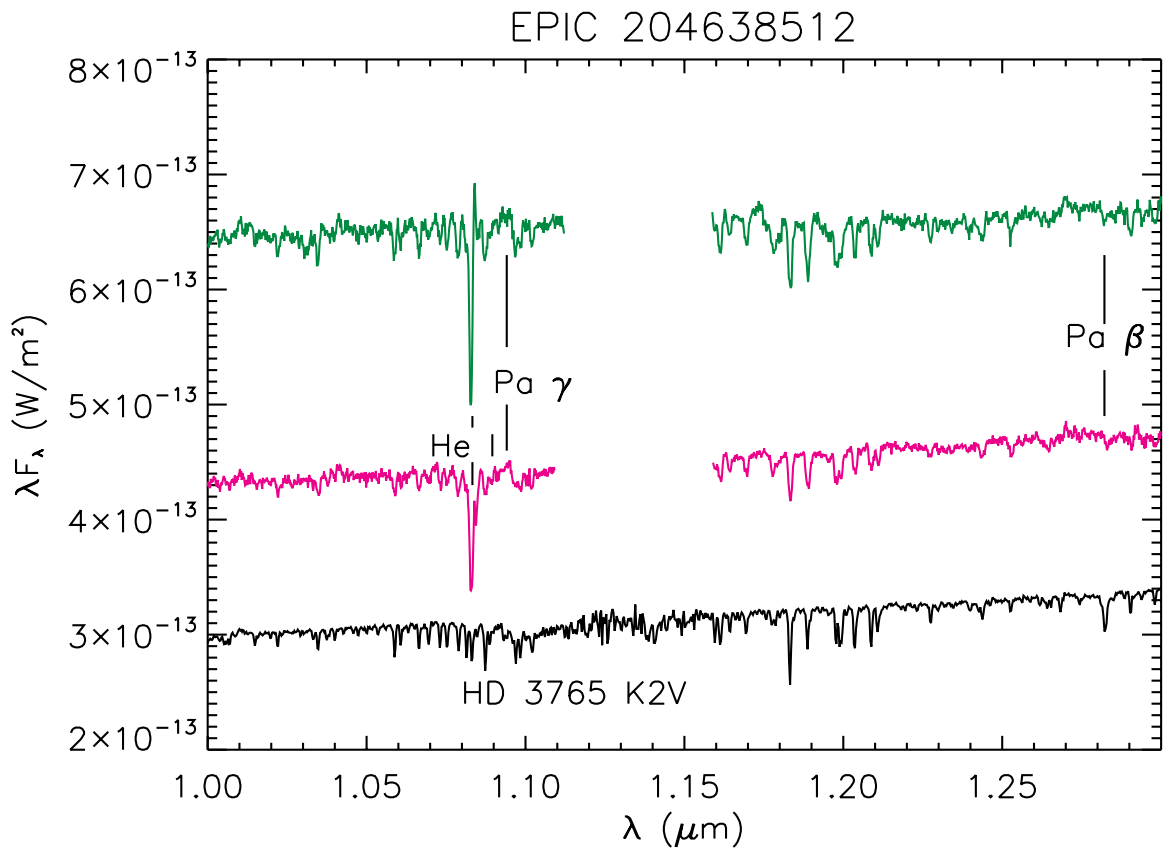}
\caption{The region near He I line for EPIC 204638512 on two consecutive nights, along with the spectrum of HD 3765, a star of similar spectral type, taken from the SpeX Atlas of Cool Stars \citep{rayner09}. The gap in the data on EPIC 204628512 is the result of trimming out the spectrum heavily affected by telluric absorption. Also shown are the locations of two nearby hydrogen lines, often seen in emission in stars actively accreting material from their disks, and the He I line. The Pa $\beta$ line is weaker than in HD 3765, suggesting it might be weakly in emission, partially filling the photospheric absorption line (in K2 V stars this region also contains a blend of Ti I, Ni I, Fe I, and Ca I lines - see Fig. 30 of \citet{rayner09}). In the brighter continuum state, the He I line exhibits a weak emission component in addition to a strong absorption core at shorter wavelengths. This resembles a P Cygni-like profile,  but with a relatively stronger short wavelength absorption component. In the higher dust occultation data, the red wing of the He I line is not evident. \label{fig:f4}}
\end{figure}

\begin{figure}
\plotone{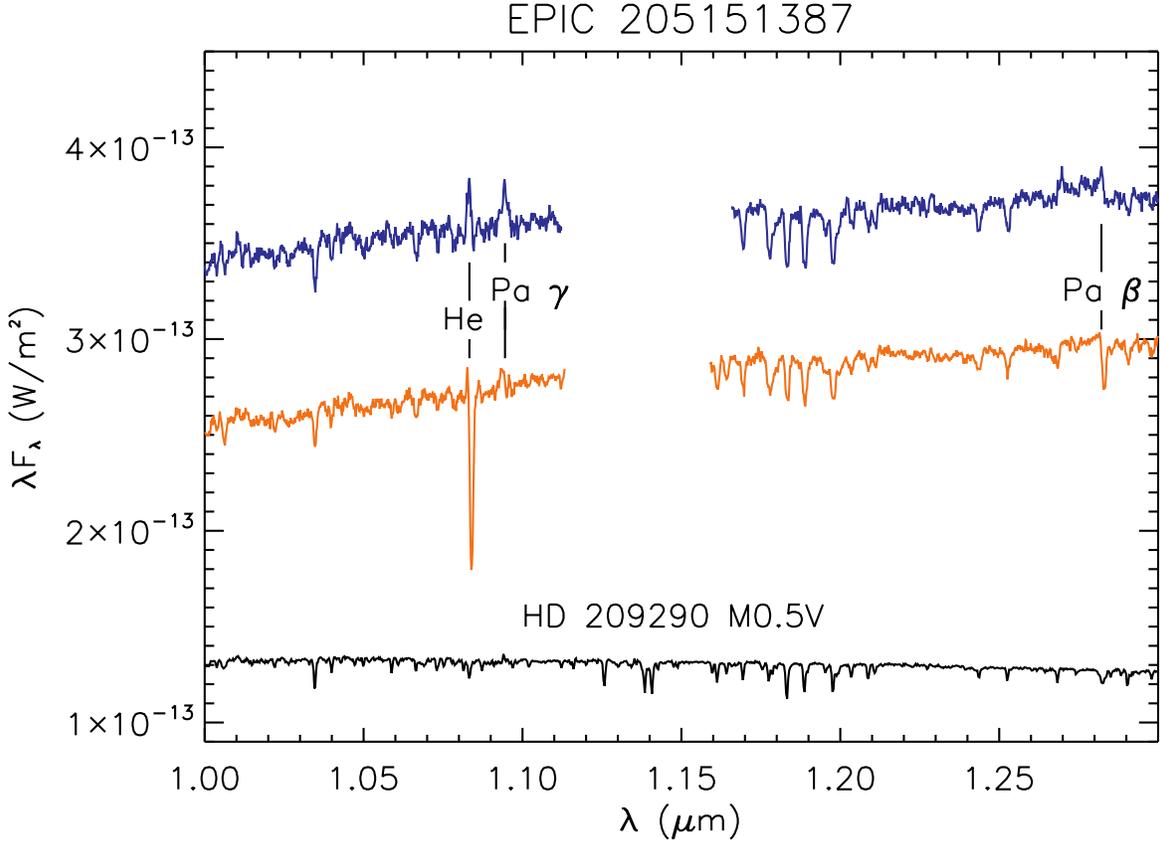}
\caption{The region near He I line for EPIC 205151387 on two consecutive nights, along with the spectrum of HD 209290, a star of similar spectral type, taken from the SpeX Atlas of Cool Stars \citep{rayner09}. During the dust extinction event (lower net flux) , the He I absorption line became even stronger than in EPIC 204628512, The Pa $\beta$ line developed a distinct absorption component, and the emission on Pa $\gamma$ was weakened. The absorption in He I and Pa $\beta$ are both shifted toward longer wavelengths than the emission, inverse P Cygni-like profiles, indicating infalling gas. \label{fig:f5}}
\end{figure}

\begin{figure}
\plotone{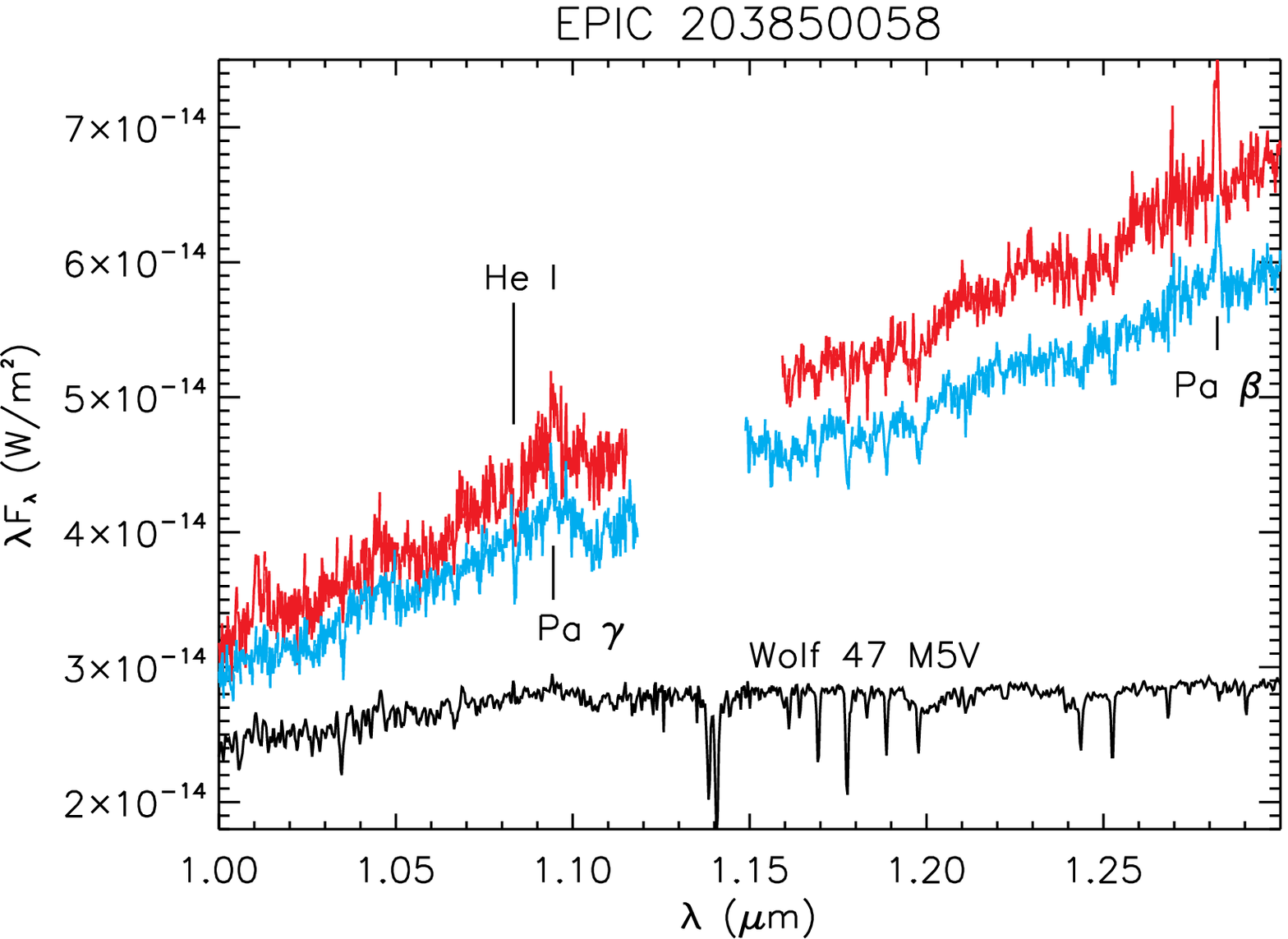}
\caption{The region near He I line for EPIC 203850058 on two consecutive nights, along with the spectrum of Wolf 47, a star of similar spectral type, taken from the SpeX Atlas of Cool Stars \citep{rayner09}. The Pa $\beta$ and Pa $\gamma$ lines are in emission, suggesting some accreting gas. The He I line exhibits absorption, but unlike EPIC 204638512 and EPIN 501551387, does not show a detectable He I emission component, although the quality of the spectra are inferior to those of the other two dippers. Note that Wolf 47 is a dMe star exhibiting flares (including at radio wavelengths \citet{white89}) and so some line emission (Pa $\gamma$?) is not unexpected. \label{fig:f6}}
\end{figure}

For completeness, we also show the two spectra of EPIC 203850058 (Fig. 5), but the data were of insufficient quality for an analysis of the extinction.

\section{Discussion}

Using ALMA dust continuum and CO line emission data of EPIS 204638512, \citet{mayama18} have provided convincing evidence of an inner disk that is misaligned from the outer disk, with an inclination of -45$^\circ$. \citet{pinilla18} have also found that the shadows cast by the inner disk onto the outer disk are variable in time. These results clearly indicate that the ``dipper" phenomenon EPIC 2064638512 are due to dust in the inclined inner disk producing variable extinction by the inclined inner disk. While less-well studied, it suggests that the extinction events seen in EPIC 205251387 have a similar origin. 

In this paper we have derived the extinction properties of the dust in these twi sources that imply grains that are unlike those found in the general ISM from which they originally formed.  Rather, the occulting grains are those that at some earlier epoch have undergone significant growth in size. Such grain growth is expected in disks \citep{dalessio01,dalessio06,furlan05}, and is generally thought to be accompanied by settling of the larger grains to the disk mid-plane, as they begin to decouple from the gas. In the case of EPIC 205151387 and EPIC 204638512, the types of grains present are consistent with the sorts of grains expected in disk systems. Indeed, the extinction properties of the grains used were \textit{derived} from the scattered light and thermal emission of the disk of HH30 \citep{cotera01, wood02}. However, if the dust particles causing the Dipper events are this large, then the ``settled dust" dominates these extinction events. In this case we may be seeing one possible near-final evolutionary stage of a disk system. The structure of the CO maps in EPIC 204638512 might also hint at infalling material \citep{mayama18}, but the evidence is inconclusive. The evidence seems stronger that this system possesses a mis-inclined inner disk that produces variable amounts of shadowing nearly 140-180$^\circ$. apart in azimuth in the outer disk. ``Dipper'' occultation events would be more likely than if the inner regions had the low inclinations originally reported for the outer disk \citep{ansdell16b}.\\

While the ``Dipper" events are almost inevitable for highly inclined inner disks, is there evidence for similar structure in disks not fortuitously observed at high inclination? A number of disks exhibit outer disk shadows that imply mis-aligned inner disks: HD 142527 \citep{avenhaus14}, HD 100453 \citep{benisty17, long17}, TW Hya \citep{debes17,poteet18}, and HD 100453 \citep{garufi16}. A related type of phenomenon might also be occurring in the non-Dipper star SAO 206462 (HD 135344B). Recent SPHERE imaging of its disk by \citet{stolker17} showed \textit{moving shadows} in the outer disk. These traveled many degrees in the course of months, suggesting varying shadows by irregular dust clouds in the inner disk. This apparent orbital motion was accompanied by changes in color of the disk in general, suggesting that the outer disk was experiencing changing illumination due to clouds that effectively reddened the light reaching it. Were its inner disk highly inclined with respect to the viewer, it might exhibit ``Dipper"-like events. Careful monitoring of near-IR spectra of systems with evidence for mis-aligned inner disks should be pursued. \\

\section{Conclusions}

We have found that the ``dipper" events observed in EPIC 20435638512 are due to changes in the extinction by dust grains whose sizes indicate significant grain growth compared to their ISM sizes, and these are located in the surface layers of an inner disk that is highly inclined to the outer disk, seen in ALMA continuum images.  Our results also suggest that other Dippers having outer disks with modest inclinations (such as EPIC 205151387, with $i=53\pm2^\circ$ might contain misaligned inner disks. The detection of shadows, especially those with wide position angle separations and/or variable shadows in the outer disk would help confirm such an hypothesis, as would interferometric observations of the inner regions.\\

As to the gas velocity structures in EPIC 204638512 and EPIC 205151387, the situation is complex. Both outflowing (wind) and inflowing (accretion) gas is present in differing degrees at different epochs. Further work that includes a number of emission/absorption lines simultaneous with thermal dust extinction measurements are needed.

This paper was based on observations taken with the IRTF/SpeX 0.8-5.5 Micron Medium-Resolution Spectrograph and Imager, funded by the National Science Foundation and NASA and operated by he NASA Infrared Telescope Facility.The authors would like to thank the tireless support of of the staff of NASA's Infrared Telescope Facility.  We wish to emphasize the pivotal cultural role and reverence that the summit of Maunakea has always had within the indigenous Hawaiian community. We are most fortunate to have the privilege to conduct scientific observations from this mountain. This work was supported in part by NASA XRP grants NNX17AF88G and NNX16AJ75G (MLS) and 80NSSC20KO254 (CAG)..

\software{HOCHUNK3D (Whitney et al. 2013, ApJS, 207, 30), Spextool(Vacca et al. 2003, PASP, 115, 389; Cushing et al. 2004, PASP, 116, 362)}

\clearpage

\clearpage

\end{document}